\documentclass[structabstract]{aa}  
\usepackage{graphicx}
\usepackage{txfonts}
\usepackage{natbib}
\bibpunct{(}{)}{;}{a}{}{,} 
\usepackage{color}
\usepackage{lscape}

\begin{document}
   \title{The Clump Mass Function of the Dense Clouds in the Carina Nebula Complex
   \thanks{Based on data acquired with the Atacama Pathfinder Experiment (APEX). 
   APEX is a collaboration between the Max-Planck-Institut f\"ur Radioastronomie, 
   the European Southern Observatory, and the Onsala Space Observatory.}}

   \titlerunning{ClMF of Dense Clouds in the Carina Nebula Complex}
   \authorrunning{S. Pekruhl et al.}
   \author{Stephanie Pekruhl \inst{1}
          \and
          Thomas Preibisch \inst{1}
         \and
         Frederic Schuller \inst{2}
         \and
         Karl Menten \inst{2}
                    }

   \institute{Universit\"ats-Sternwarte M\"unchen,
              Ludwig-Maximilians-Universit\"at,
              Scheinerstr.~1, 81679 M\"unchen, Germany\\
              \email{pekruhl@usm.uni-muenchen.de}
        \and
            Max-Planck-Institut f\"ur Radioastronomie,
            Auf dem H\"ugel 69, 53121 Bonn, Germany
             }

   \date{}

 
  \abstract
   {The question how the initial conditions in a star-forming region
    affect the resulting mass function of the forming stars is one of
    the most fundamental open topics in star formation theory.}
   {We want to characterize the properties of the cold dust clumps in
    the Carina Nebula Complex, which is one of the most massive star
    forming regions in our Galaxy and shows a very high level of
    massive star feedback. We derive the Clump Mass Function (ClMF),
    explore the reliability of different clump extraction algorithms,
    and investigate the influence of the temperatures within the
    clouds on the resulting shape of the ClMF.}
   {We analyze a $1.25\hbox{$^\circ$} \times 1.25\hbox{$^\circ$}$
    wide-field submillimeter map obtained with LABOCA at the APEX
    telescope, which provides the first spatially complete survey of
    the clouds in the Carina Nebula Complex. We use the three 
    clump-finding algorithms CLUMPFIND, GAUSSCLUMPS and SExtractor to
    identify individual clumps and determine their total fluxes.
    In addition to assuming a common ``typical'' temperature for all
    clouds, we also employ an empirical relation between cloud column
    densities and temperature to determine an estimate of the 
    individual clump temperatures, and use this to determine 
    individual clump masses.}
   {We find that the ClMFs resulting from the different extraction
    methods show considerable differences in their shape. While the
    ClMF based on the CLUMPFIND extraction is very well described by
    a power-law (for clump masses well above the completeness limit), 
    the ClMFs based on the extractions with GAUSSCLUMPS and 
    SExtractor are better represented by a log-normal distribution. 
    We also find that the use of individual clump temperatures leads 
    to a shallower ClMF slope than the (often used) assumption of a
    common temperature (e.g.~20~K) of all clumps.}
   {The power-law of $dN/dM \propto M^{-1.95}$ we find for the
    CLUMPFIND sample is in good agreement with ClMF slopes found in
    previous studies of the ClMFs of other regions. The dependence of
    the ClMF shape (power-law versus log-normal distribution) on the 
    employed extraction method suggests that observational
    determinations of the ClMF shape yields only very limited 
    information about the true structure of the cloud. 
    Interpretations of log-normal ClMF shape as a signature of 
    turbulent pre-stellar clouds versus power-law ClMFs as a
    signature of star-forming clouds may be taken with caution for a
    single extraction algorithm without additional information.}


   \maketitle
%

\section{Introduction}

The process of star formation occurs in a wide variety of
environments and corresponding physical conditions. Most nearby
($d \la 300$~pc) star-forming regions are low-density clusters or
associations in which only low- and intermediate mass stars form.The 
interaction between the young stars in such regions is minimal and 
they can thus be considered as forming essentially in isolation. The 
more massive and generally more distant star-forming regions, on the 
other hand, contain high-mass stars ($M \ga 20\,M_\odot$), too. These 
massive stars profoundly influence their environments by creating HII 
regions, generating wind-blown bubbles, and, finally, exploding as 
supernovae. This massive star feedback can disperse the natal 
molecular clouds, but ionization fronts and expanding superbubbles 
can also compress nearby clouds and may thereby trigger the formation 
of new generations of stars.

Despite these differences in the formation environment, the mass
function of the forming stars, i.e.~the final result of the star
formation process, appears to be remarkably uniform \citep[see][and
references therein]{Bastian10}. \citet{Salpeter55} showed that the
distribution of initial stellar masses can be described by a
power-law $dN/dM \propto M^{-\alpha}$ with $\alpha=2.35$. Further 
studies confirmed the power-law shape of the Initial Mass Function 
(IMF) with a power-law index $\alpha$ between 2.1 and 2.5 in the 
upper mass regime (above $\sim 1\,M_\odot$ ) 
\citep{Kroupa01,Schneider04,Andre10} and showed a shallower slope and 
a turn-over for lower masses. Alternatively, the IMF can be described 
by a log-normal distribution \citep{Chabrier03}.

One of the most fundamental open questions in star formation theory 
is how the stellar IMF is related to the initial molecular cloud 
density structure. Submillimeter observations have become an 
important tool for investigating these dense cloud structures. LABOCA 
has already been used for surveys of distant molecular clouds and 
clumps \citep{Schuller09, Bot10}.

In observations of the generally distant massive star-forming 
regions, the individual cloud cores can usually not be resolved at 
submillimeter or radio wavelengths. The relevant structures 
accessible to the observations are then the \textit{clumps} within a 
cloud and the corresponding Clump Mass Function (ClMF). Molecular 
line \citep{Kramer98, Wong08} and dust continuum emission 
\citep{Johnstone06, Munoz07} observations of several star-forming
regions show that the ClMF of molecular clouds also can be described 
by a power-law distribution. The slope of the ClMF, with $\alpha \sim 
1.4-2.0$ \citep{Elmegreen96}, is, however, typically shallower than 
the core mass function and the stellar IMF.

Recent studies of molecular cloud structure suggest that the ClMF or
the distribution function of the column density (N-PDF) can be used 
as an indicator for the evolutionary state of a molecular cloud
\citep{Kainulainen09, Kainulainen11, Ballesteros11}. In clouds in 
which the star formation process has not yet started, turbulence is 
expected to lead to a log-normal distribution. As soon as the  star 
formation process starts, the denser structures get dominated by 
gravity and this should result in a power-law distribution of the 
masses (or column densities) in the upper mass (density) range. 
Therefore, the shape of the observed ClMF or density distribution 
function is sometimes considered as an indicator for the physical 
status of a cloud. One problem with such an interpretation is that 
different observing and analysis techniques sometimes yield different 
ClMF shapes for the same cloud \citep[see][]{Reid06, Reid10}. Further 
determinations of the ClMF of clouds in star-forming regions are 
required to gain more insight into these questions.

The Carina Nebula Complex represents one of the most massive Galactic
star-forming regions \cite[see][for an overview]{Smith08}. It is
located in the Sagittarius-Carina spiral arm at a very well known
distance of 2.3 kpc (Smith~2002), and hosts at least 65 O stars
\citep[including several O3 stars and the O2 supergiant HD~93129A; 
see][]{Smith07}, several Wolf-Rayet stars, and the Luminous Blue 
Variable $\eta$ Carinae, which is our Galaxy's most luminous star 
known and expected to explode as a supernova within the next Myr. 
These massive stars are located in several loose clusters 
\citep[Tr~14, Tr~15, Tr~16, see][]{Trumpler30}, with ages ranging 
from around 1 to several Myr. The feedback from the massive stars has 
already dispersed most of the initial clouds in the central region. 
The radiation and stellar winds of the massive stars have formed 
numerous giant dust pillars \citep[South Pillars, see][]{Smith06} 
located a few parsecs away from the stars in south-eastern and 
north-western direction. Several studies found clear indications for  
ongoing and triggered star formation in the tips of these pillars 
\citep{Megeath96,Smith08, Smith10b}. 

The stellar populations in the Carina Nebula Complex have recently 
been studied in detail in the context of comprehensive multi-
wavelength surveys. A deep wide-field {\it Chandra} X-ray surveys has 
allowed to detect a large sample of the low-mass stellar population
\citep{Townsley11,Preibisch11a,Wang11,Wolk11}. A very deep 
near-infrared survey of the central region with HAWK-I at the ESO 
Very Large Telescope allowed a detailed characterization of the 
properties of the X-ray selected young stars \citep{Preibisch11b}. 
Additional information about the recent star formation processes came 
from {\it Hubble} Space Telescope (HST) observations of protostellar 
jets \citep{Smith10a} and {\it Spitzer} infrared imaging of the South
Pillars region \citep{Smith10b}.

While the stellar populations are by now rather well investigated, 
the available observations of the molecular clouds (and thus on the 
interaction of the massive stars and the surrounding clouds) were 
quite limited until recently. The most extensive existing data set of 
the Carina Nebula Complex at radio wavelengths is a NANTEN survey in 
several CO lines, covering a $4\degr \times 2\degr$ area by 
\citet{Yonekura05}. Mopra $^{12}$CO (1--0) data of a smaller part of
the central region, but with higher spatial resolution has been
presented by \citet{Brooks98} \citep[see also][]{Schneider04}.

We have recently performed a large-scale
($1.25\hbox{$^\circ$} \times 1.25\hbox{$^\circ$}$) submillimeter
mapping of the Carina Nebula Complex with LABOCA at the APEX 
telescope to obtain detailed information on the structure of the cold 
dusty clouds \citep{Preibisch11}. Our LABOCA data revealed the very 
clumpy structure of the clouds. We found that the total mass of the 
dense clouds to which LABOCA is sensitive is $\sim 60\,000\,M_\odot$. 
This value agrees fairly well with the mass estimates for the well 
localized molecular gas traced by $^{13}$CO \citep{Yonekura05}. This 
high mass emphasizes that despite several mega-years of ongoing cloud 
destruction due to massive star feedback, there is still a very large 
amount of cloud material available for future star formation.

In this paper, we analyse the distribution of the cloud masses in the
LABOCA map. We use the three common clump-finding algorithms 
CLUMPFIND \citep{Williams94}, GAUSSCLUMPS \citep{Stutzki90} and 
SExtractor \citep{Bertin96} to identify individual cloud clumps and 
to measure their total sum-mm fluxes (Sect.~2). In Sect.~3 we 
estimate individual temperatures for each clump and calculate the 
clump masses. We then discuss our results and the following 
consequences. Finally, Sect.~4 summarizes the results of this study.


\section{Observation and data analysis}

   \begin{figure*}
   \centering
   \includegraphics[width=\textwidth]{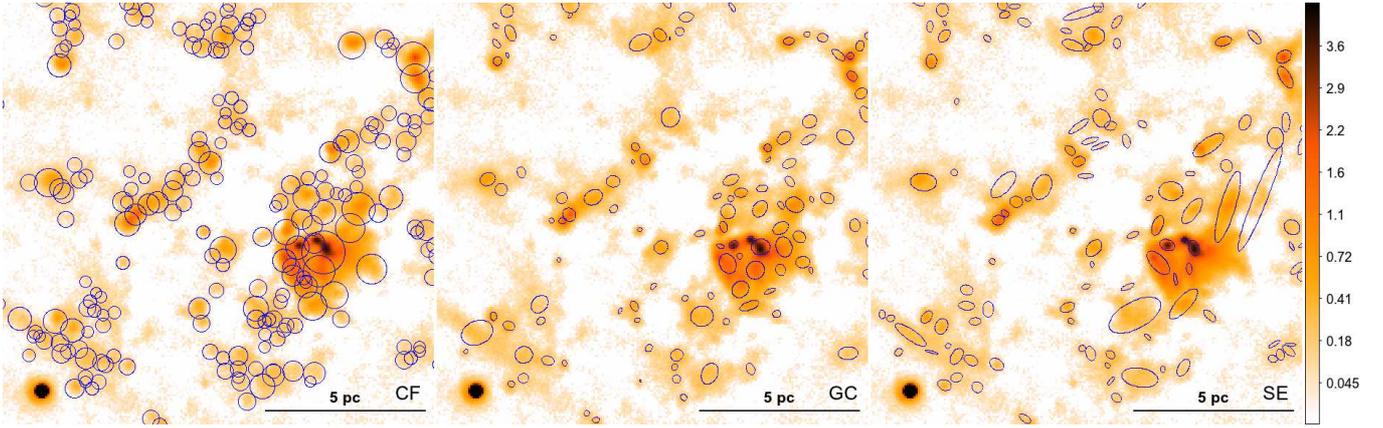} 
   \caption{A $20$~arcminutes wide detail from our LABOCA map at
   $870\,\mu$m of the Tr~14 region in the Carina Nebula Complex with
   the positions of the clumps found with the different clump-finding
   algorithms. The units of the scale bar on the right are Jy/beam.
   In the left panel the CLUMPFIND sample is shown with circles of
   the effective radial size of the clumps calculated by the 
   algorithm. The GAUSSCLUMPS (middle panel) and SExtractor (right
   panel) samples are shown as ellipses. The sizes of their axes
   correspond to the rms deviations.
   The bright source in the lower left corner is the luminous blue
   variable $\eta$~Car.}
   \label{sample}
   \end{figure*}

\subsection{Observations}
To investigate the clumpy dust structures and to determine the ClMF 
in the Carina Nebula we used the sub-mm emission data of the
$1.25\hbox{$^\circ$} \times 1.25\hbox{$^\circ$}$ wide-field map we
derived from observations we performed at the 12-meter APEX
\citep[''Atacama Pathfinder Experiment'',][]{Guesten06} telescope,
with the Large Apex BOlometer CAmera LABOCA \citep[see][]{Siringo09},
which operates in the atmospheric window at $870\,\mu$m. LABOCA has 
an angular resolution of $18.6''$ which, at the distance of the 
Carina Nebula, allows us to resolve structures down to 0.2~pc. The 
map thus is suitable to resolve the molecular clumps, but not the 
individual cores and provides the first spatially complete survey of 
the dust clouds in the Carina Nebula. These clumps are expected to be 
the sites at which star clusters can form, contrary to prestellar 
cores which are supposed to form single or gravitationally bound 
multiple protostars \citep{Williams00}. 

For the data reduction the BOlometer array Analysis software (BOA)
\citep{Schuller09} has been used, resulting in a pixel size of 
$6.07''$ in the final map. The surface brightness can be transformed 
to integrated fluxes by multiplying it with the pixel-to-beam-size 
ratio (0.0941 beams/pixel). The average rms~noise level for the map 
is about 20~mJy/beam, which corresponds for isolated compact clumps 
with estimated temperatures of $T \approx 20-30$~K to a sensitivity 
limit in mass of about $2\,M_\odot$. For the clouds we measure 
intensities up to around 4~Jy/beam. The total flux measured in the 
map above a $3\sigma$ noise level amounts 1147~Jy. The observations 
are described in detail in \citet{Preibisch11}.

\subsection{Clump-finding algorithms}
\label{Clump-finding algorithms}

   \begin{figure}
   \centering
   \resizebox{\hsize}{!}{\includegraphics{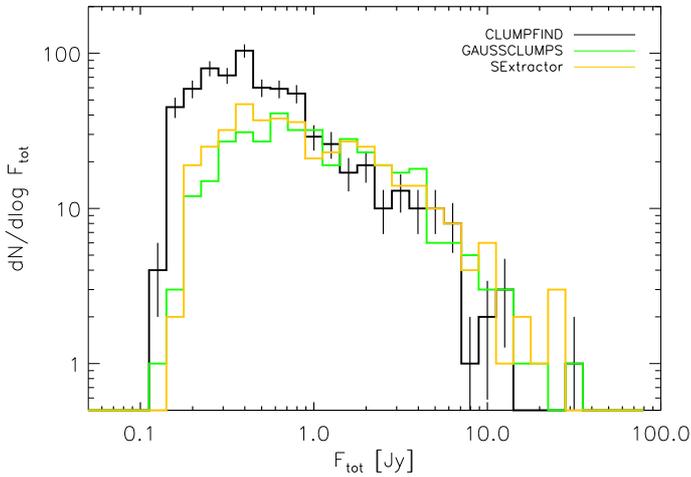}}
   \caption{The flux distribution of clumps for the three different
    clump-finding algorithms with a statistical error
    $\sqrt{\frac{dN}{d\log F}}$ for the CLUMPFIND sample.}
   \label{flux_comb}
   \end{figure}

\subsubsection{CLUMPFIND}
\label{CLUMPFIND}
The CLUMPFIND (CF) algorithm from \citet{Williams94} contours the 
data at a given threshold, which is recommended to be chosen as three 
times the rms~noise of the map. Afterwards it identifies the peaks of
emission which locate the clumps. It then contours the data in 
discrete steps defined by the user down to lower intensities until 
the lowest threshold is reached, assigning the pixels within the 
contours to the before located maxima. Blended structures are 
assigned via a friends-to-friends algorithm to the closest peak. No 
specific clump profile is assumed.

As we find an rms~noise level of about 20~mJy/beam for our data we
start contouring at an intensity level of $60$~mJy/beam and go in 
steps of $40$~mJy/beam up to $4.5$~Jy/beam, below which all the  
emission of the clouds is found. CLUMPFIND extracts 687 molecular 
clumps with a total flux of 720~Jy from the LABOCA map. This is about 
63\% of the total flux in the map. 
A section of the central region around the star cluster Tr~14, with 
the clumps detected by CLUMPFIND marked as blue circles, is shown in 
the left panel of Fig.~\ref{sample}. The strong point-like source 
$\eta$~Car has been excluded from this and the following samples. 
Fig.~\ref{flux_comb} shows the distribution of the extracted total 
fluxes $F_{tot}$.

\subsubsection{GAUSSCLUMPS}
\label{GAUSSCLUMPS}
GAUSSCLUMPS (GC) \citep{Stutzki90} is a least square fitting 
algorithm, that assumes the clumps to have a Gaussian shape and works 
directly on the continuous intensity distribution of the data. It 
first searches for the highest intensity peak on the data map and 
fits a Gaussian shaped clump around it. Then the algorithm subtracts 
the clump from the original map generating a new map on which it 
continues with the next iteration step.

The GAUSSCLUMPS algorithm calculates an rms~noise level of 
22~mJy/beam, and we define a detection threshold of $3\,\sigma$, so 
we can compare the results. 371~clumps are detected, only about half 
the number found by CLUMPFIND (see Fig.~\ref{sample}, middle panel;
Fig.~\ref{flux_comb}). Nevertheless, the total flux of 728~Jy
within these clumps, is nearly the same as for the CLUMPFIND clumps.

\subsubsection{SExtractor}
\label{SExtractor}
The third algorithm we used to analyse the data is the SExtractor 
(SE) software from \citet{Bertin96}, which originally was developed 
to detect stars and galaxies in large surveys. However the algorithm 
has also been used to extract sources in sub-mm observations of 
molecular cloud structures \citep{Schuller09, Coppin00}. Like the 
CLUMPFIND algorithm it uses the thresholding method to locate the 
clumps. To deblend sources the algorithm contours them at 30 
exponentially spaced levels between the peak values of the detected 
sources and the detection threshold and follows the structure 
downwards. At each of this levels it tests if there is another peak, 
with an intensity exceeding a certain fraction of the total 
intensity. If this is the case, it extracts a new source. For each 
pixel between maxima, SExtractor calculates its contribution to each 
object by assuming a Gaussian profile, and converts this into a 
probability for that pixel to be assigned to a certain object. 

SExtractor determined an rms noise~level of about 30~mJy/beam so we 
used a detection threshold of $2 \sigma$ to make our results 
comparable to the other algorithms. Corresponding to our resolution 
we defined the minimal area of a detected clump to 10~pixel. 
SExtractor finds 432~sources (see Fig.~\ref{sample}, right panel; 
Fig.~\ref{flux_comb}) with a total flux of 823~Jy, 72\% of the total 
flux in the map. This is slightly higher than the results of the 
other algorithms.

All the extracted sources of the SExtractor sample, as well as the 
clumps identified by the former algorithms, have fluxes above 
$1.6$~Jy/beam, except 18 of them that we considered artefacts and do 
not further take into account in our analysis.

\subsection{Common Sources}

   \begin{figure}
   \centering
   \resizebox{\hsize}{!}{\includegraphics{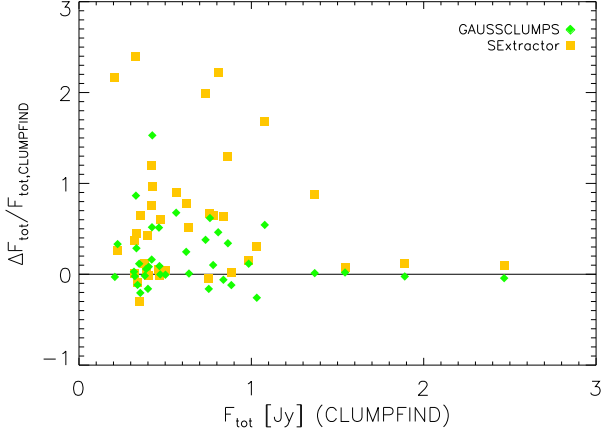}}
   \caption{The relative difference of the GAUSSCLUMPS 
   (green) and SExtractor (yellow) fluxes with respect to the
   CLUMPFIND fluxes of the 38 commonly identified clumps.
}
   \label{flux_tot}
   \end{figure}

The three used extraction algorithms are working along different 
principles. While CLUMPFIND assumes no specific clump profile, the 
other two algorithms presume a Gaussian shape. Furthermore the 
CLUMPFIND and SExtractor algorithms use thresholding methods, while 
the GAUSSCLUMPS algorithm works on residual maps.

In order to check how well the extraction results of the different
algorithms agree, we defined a sample of ``common sources'', where
the same clump is detected by all three algorithms. In the crowded 
and filamentary clouds in the center of the Carina nebula, many 
clumps are blended; therefore, the different algorithms produce often 
quite different decomposition results.

However, in the less crowded parts of our map, we could identify 38 
rather isolated clumps which are detected by all three algorithms in 
a consistent way. For these, we compare the derived total fluxes in 
Fig.~3, where we plot the relative difference of the GAUSSCLUMPS and 
SExtractor fluxes with respect to the CLUMPFIND fluxes. The plot 
shows a generally reasonable agreement, although differences by 
factors up to $\sim 2-3$ are seen in the total fluxes for some of the 
faintest clumps in the common sample. For the majority of the clumps 
the deviations are less than 50\%, and for most of the brighter 
clumps very good agreement, with deviations of less than 10\%, is 
found. From this we conclude that for sufficiently bright isolated 
clumps the three extraction algorithms yield consistent total fluxes.

\section{Results and Discussion}

\subsection{Determination of clump temperatures and masses}

The masses of clumps can be calculated from the observed sub-mm flux 
of thermal dust emission, which is optically thin, via

\begin{equation}
M = \frac{d^2\,F_{\nu,\, \mathrm{tot}} \,R}{B_{\nu}(T_d)\;\kappa_{\nu}}\;\; ,
\end{equation}

\noindent
where $F_{\nu,\, \mathrm{tot}}$ is the observed spectral flux 
integrated over the source and $d$ is the distance of the molecular 
cloud. For the gas-to-dust mass ratio $R$ and the dust emissivity 
$\kappa_{\nu}$ we used the values from \citet{Schuller09} i.e. 
$R=100$ and 
$\kappa_{870\,\mu{\rm m}} = 1.85\,{\rm cm}^2\,{\rm g}^{-1}$.
$B_{\nu}(T_d)$ is the Planck function for a given dust temperature 
$T_d$ \citep[see also][]{Preibisch11}.

A difference in temperature of only a few degrees in the relevant 
regime can change the derived masses of the clumps by a factor of 
$2-3$ and therefore strongly affect the derived ClMF of the region
\citep{Stamatellos07}. Hence, for a reliable mass estimation a good 
approximation of the temperatures of the clumps is indispensable.

Unfortunately, for single-wavelength data sets (such as our LABOCA 
map) no direct determination of the cloud temperature is possible. In 
many investigations it is assumed that all clouds in a map would have 
a common temperature, e.g.~15~K \citep[e.g.][]{Mookerjea04, Kirk06}.
While this may be a reasonable approximation in the case of more or
less isolated clouds in a quiescent environment, we believe that such
an assumption is not valid in the case of the Carina Nebula. The 
clumps in different parts of the complex will be affected by strongly
different levels of irradiation, depending on the physical distance 
from the nearest massive stars, and this should lead to considerable
differences of the cloud temperatures.

In general, clouds with high column density should be cooler than
low-column density clouds, because their interior is better shielded
from the external radiation field. Such a relation between column
density and cloud temperature has been observed in several cases, and
can provide us with important information for an estimation of
cloud masses. \citet{Peretto10} recently determined the temperatures
and the column densities of a sample of 22 infrared dark clouds 
(IRDCs) from the Hi-GAL Galactic Plane survey \citep[see][]
{Molinari10} by analyzing their spectral energy distributions. Their 
Fig.~6 shows the expected anti-correlation between column density and 
temperature. To quantify this relation, we used their column density 
peak data, performed an Ordinary Least Squares (OLS) Bisector 
fit \citep{Isobe90} and found the relation

\begin{equation}
\log N_{\rm H_2} [{\rm cm^{-2}}]=25.6-0.22\,\log T_d [{\rm K}].
\label{eq:peretto_fit}
\end{equation}

Since the column density of a clump with a dust temperature $T_d$ is
given by

\begin{equation}
N_{\rm H_2} = \frac{F_{\nu ,\, \mathrm{max}}\,R}{B_{\nu}(T_d)\;\Omega\;\kappa_{\nu}\;\mu\,m_{\rm H}}\;\; ,
\label{eq:density}
\end{equation}

\noindent
where $F_{\nu,\, \mathrm{max}}$ is the spectral peak flux, $\Omega$ 
is the beam solid angle, and $\mu$ is the mean molecular weight, we 
have two equations relating the observed flux, column density, and
temperature. We can now solve for the temperature for which the 
column density computed from Eq.~(\ref{eq:density}) is closest to the 
relation in Eq.~(\ref{eq:peretto_fit}).

   \begin{figure}
   \centering
   \includegraphics[width=0.32\textwidth]{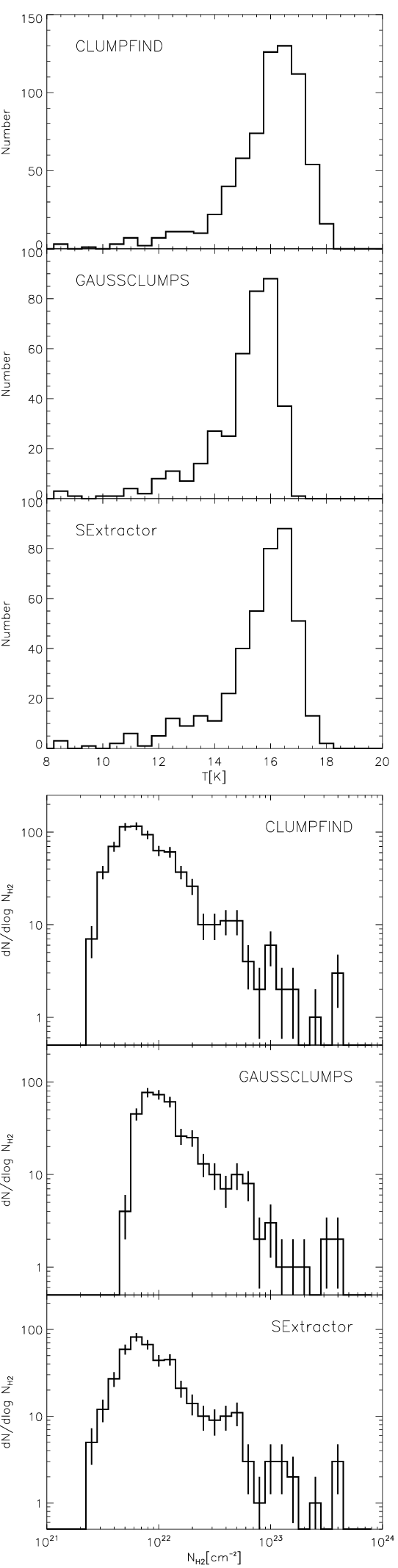}
   \caption{The distribution of individual clump temperatures (top)
   and column densities (bottom), with statistical errors, as found
   from Eq.~(\ref{eq:density}) and Eq.~(\ref{eq:peretto_fit}) for the
   three different clump-finding algorithms.}
   \label{n_T_hist}
   \end{figure}

   \begin{figure*}
   \centering
   \includegraphics[width=\textwidth]{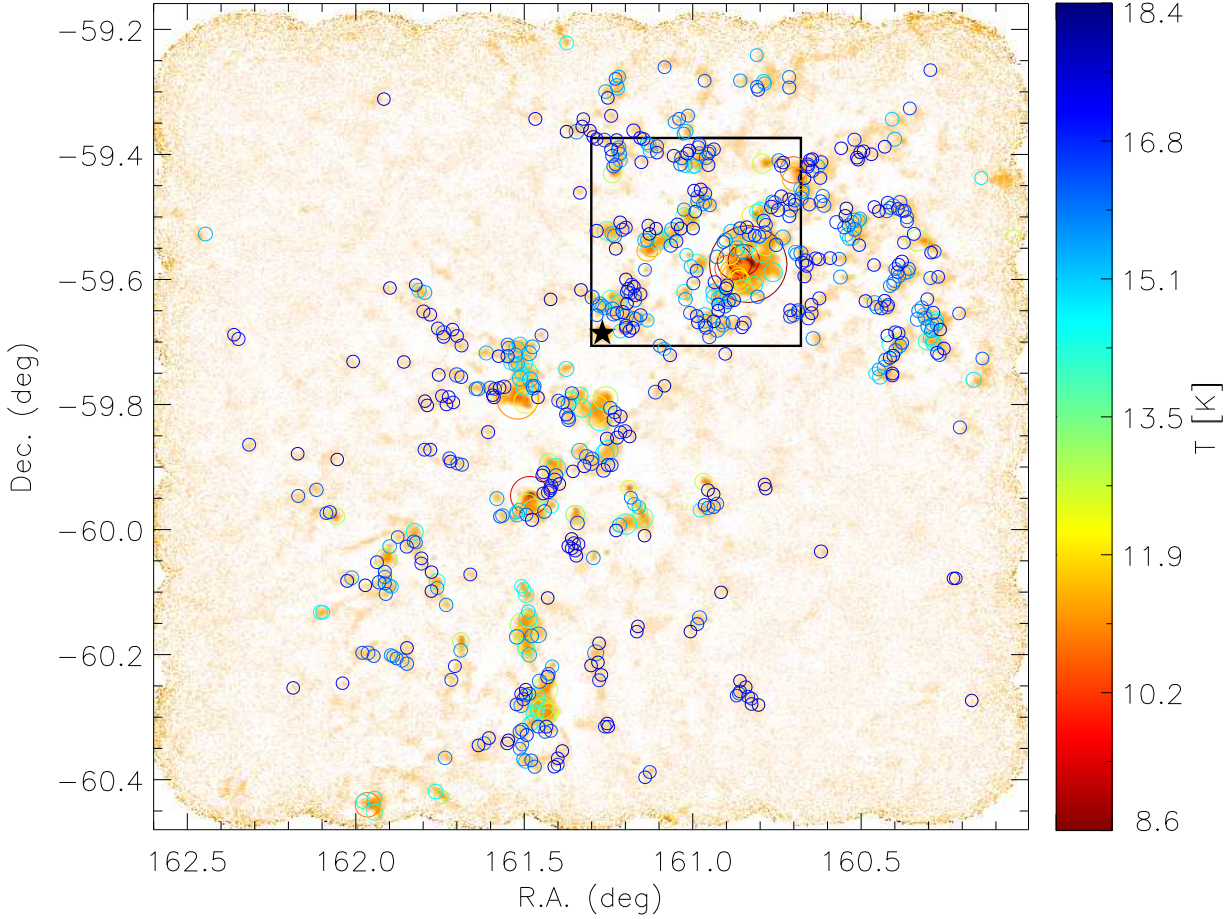}
   \caption{The temperatures and fluxes of the CLUMPFIND sample,
   overplotted on the LABOCA map. The color of the circles
   corresponds to the calculated temperatures and the size of
   the circles to the measured flux ($\sim2 - 373$~Jy/beam).
   The position of $\eta$~Car is marked by the black star. The black
   box represents the region shown in Fig.~\ref{sample}, centred
   around Tr 14 (RA(J2000)=161{\degr}, Dec~(J2000)=-59{\degr}.6).}
   \label{temp_map}
   \end{figure*}

This gives us two solutions, from which we can reject the first one 
as it leads to temperatures below 4~K. The second solution gives us, 
for all samples, temperatures between $8.5 - 18.5$~K, which are 
typical values for molecular clouds and clumps \citep{Bergin07}. Most 
of the clumps (CF: 92~\%, GC: 86~\%, SE: 87~\%) have temperatures 
above 14~K (see Fig.~\ref{n_T_hist}, top panel). The column density 
distributions we derive for the different samples are shown in the 
bottom panel of Fig.~\ref{n_T_hist}.

In Fig.~\ref{temp_map} we show the location and the temperature of 
the clumps extracted with CLUMPFIND. The clump temperatures are shown 
by the color while the size of the circles increases with the 
measured flux $F_{\nu,\, \mathrm{tot}}$ (see also 
Fig.~\ref{flux_comb}). The clumps are assembled along filamentary 
structures and pillars.
These pillars show cold dense clumps on their tips, within which star
formation occurs. Most of the brightest clumps are located within the
structures a few arcminutes to the west of the Tr~14 cluster, where 
the clouds are dense and massive. The ionizing radiation of the 
nearby cluster can only affect the cloud surfaces and does not 
pervade to the cool center of the cloud.

\subsection{The Clump Mass Function (ClMF)}

With these individual clump temperature estimates we now can derive 
the masses of individual clumps. We find masses between about 
$5\,M_\odot$ and $4.7\times10^3\,M_\odot$ and the total mass 
extracted in clumps is about $42 \times10^3\,M_\odot$ (see 
Table~\ref{Results}).

   \begin{figure}[h!]
   \centering
   \resizebox{\hsize}{!}{\includegraphics{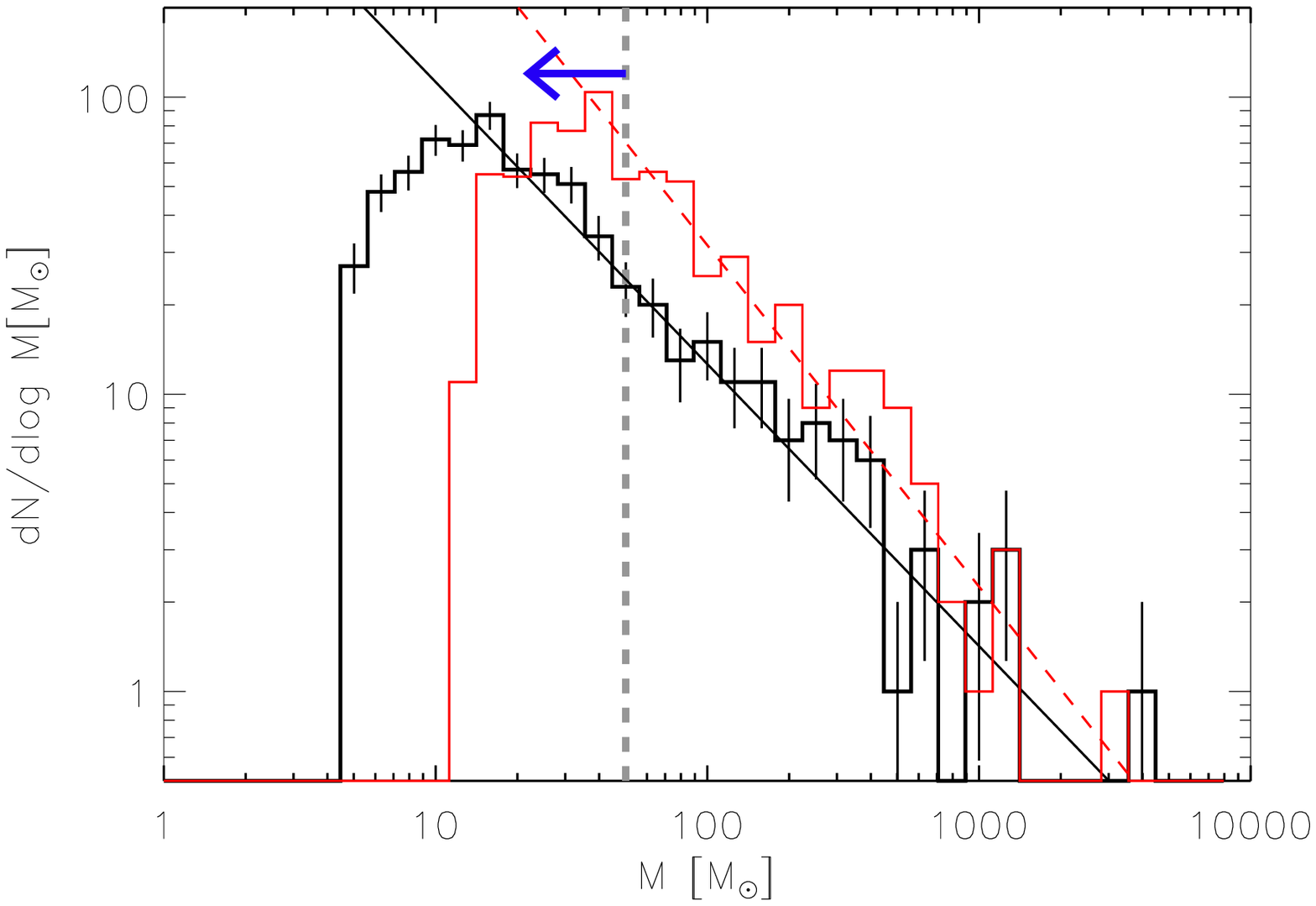}}
   \resizebox{\hsize}{!}{\includegraphics{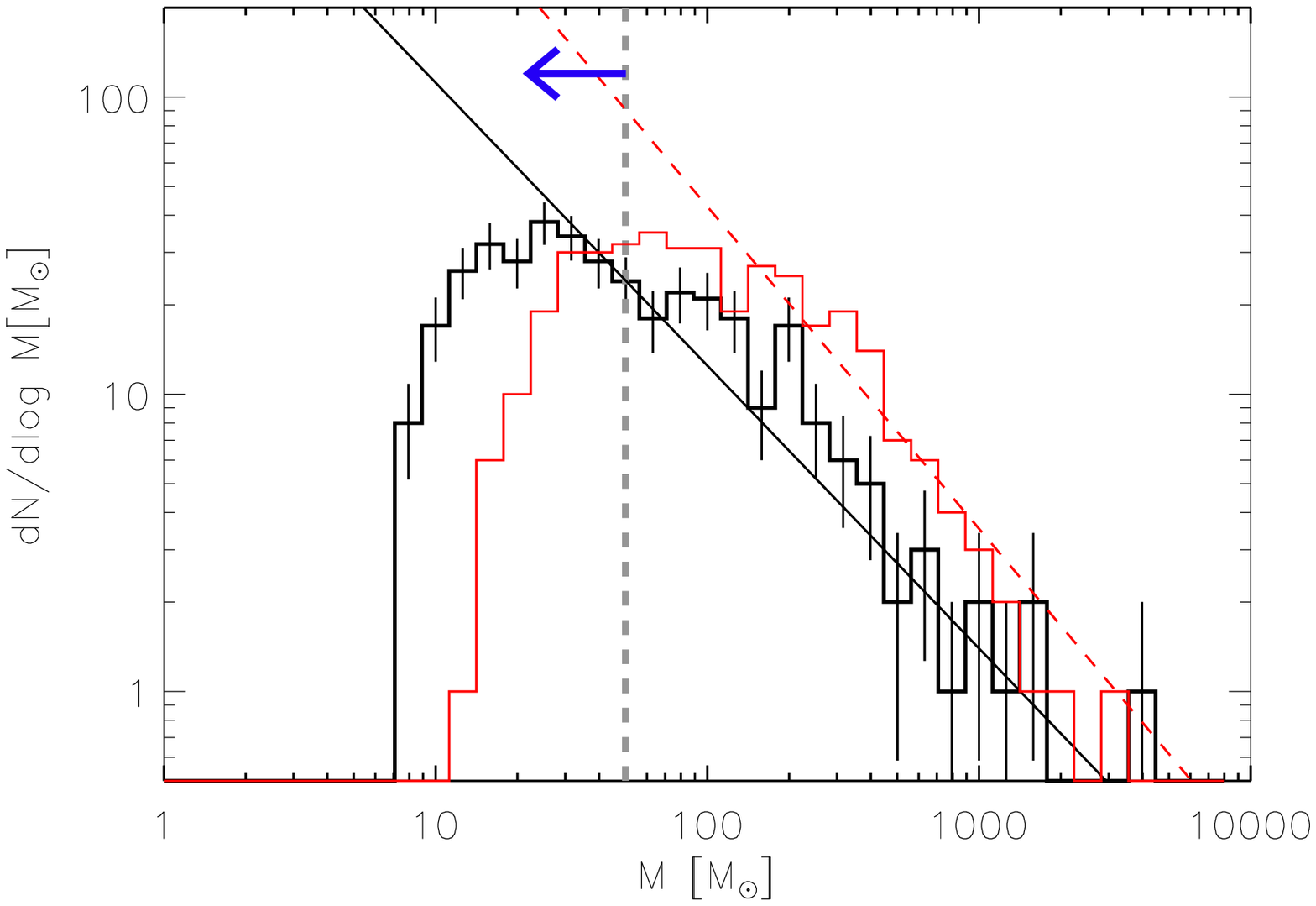}}
   \resizebox{\hsize}{!}{\includegraphics{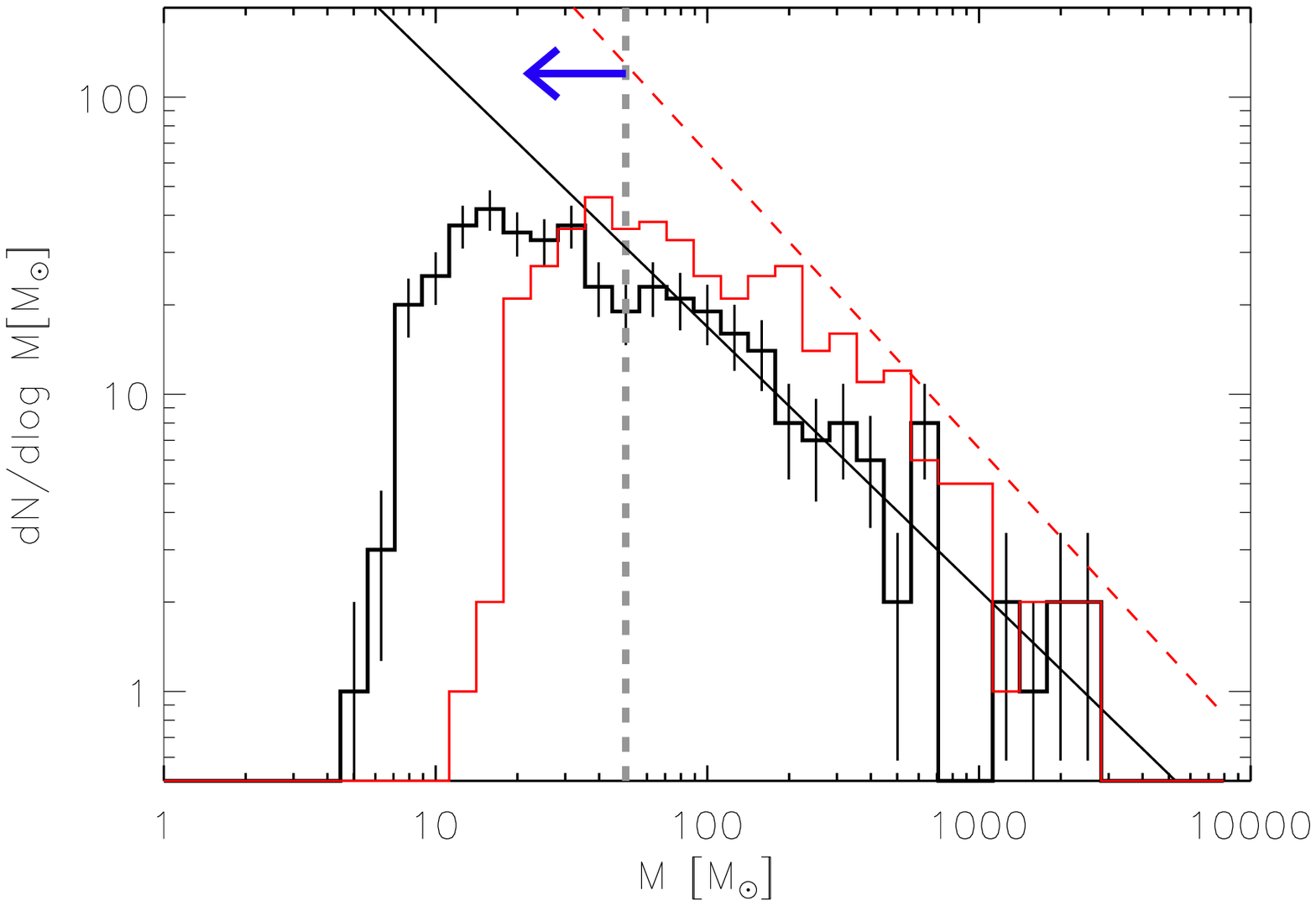}}
      \caption{The mass histograms derived with the three different
   clump-finding algorithms CLUMPFIND~(top), GAUSSCLUMPS~(middle) and
   SExtractor~(bottom). The black lines show the samples and their
   power-law slopes with the temperatures derived from the relation
   we got from Peretto et al. (2010). The dashed gray line shows the
   used lower cut-off limit for these samples. For the red line we
   assumed a constant temperature of 10~K for all clumps. If we 
   assume a constant temperature of 20~K the slope does not change,
   but the mass distribution is shifted to lower masses (blue
   arrow).}
   \label{mass_hist}
   \end{figure}

For the \textit{differential} Clump Mass Function (ClMF) we plot the
number of clumps $dN$ within a logarithmic mass interval $d\log(M)$
against the mass. Up to $\sim 10\,M_\odot$ our sample is not complete
due to the mass sensitivity limit and the wide range of background
levels at different locations in our map. For clump masses above
$\sim 10 - 20\,M_\odot$ one can see the expected decrease in the 
number of clumps for rising clump mass, which roughly follows a 
power-law

\begin{equation}
\frac{\mathrm{d}N}{\mathrm{d}M}\propto M^{-\alpha}.
\label{power-law}
\end{equation}

To derive the slope of the power-law tail of the mass spectra, we use
the method of \citet{Maschberger09} to fit the distribution. They use
a bias-free Maximum Likelihood (ML) estimator to determine the power-
law exponent without binning. This method takes into account data 
only above a given lower cut-off limit to calculate the power-law 
index. In order to make sure that the fit is not affected by 
incompleteness effects at lower masses, we use a conservative lower 
cut-off limit at $50\,M_\odot$ (gray dashed line in 
Fig.~\ref{mass_hist}) for the fit. For the power-law index we find 
$\alpha_{CF}=1.95\pm0.07$, $\alpha_{GC}=1.95\pm0.06$ and 
$\alpha_{SE}=1.89\pm0.06$ for the CLUMPFIND, GAUSSCLUMPS and  
SExtractor samples, respectively (see Table~\ref{Results}).

\subsection{ClMF assuming constant cloud temperature}

A difference in temperature of only a few degrees can change the
derived masses of the clumps significantly. Here we investigate how 
the ClMF depends on the estimated temperature. For comparison we 
therefore, also computed clump masses based on the assumption 
of isothermal clouds and considered two different values for the 
cloud temperature, 10~K and 20~K. These values are often assumed in 
sub-mm studies of molecular clouds \citep[e.g.][]{Johnstone00, 
Kirk06, Schuller09}. If we assume constant temperatures for all 
clumps the overall shape of the ClMFs keeps the same, but we derive a 
steeper power-law slope (e.g. for 10~K 
$\alpha_{CF}=2.15\pm0.08$, $\alpha_{GC}=2.08\pm0.06$ and 
$\alpha_{SE}=1.99\pm0.06$ for a lower cut-off limit of 
$100 M_\odot$). The slopes for the two different constant 
temperatures stay the same within their errors, as a change of
temperature corresponds to a constant shift in mass 
(Fig.~\ref{mass_hist}). The masses range between about $4\,M_\odot$
and $1\times10^3\,M_\odot$ for 20~K and about $14\,M_\odot$ and
$3.3\times10^3\,M_\odot$ for a temperature of 10~K. The results are
summarized in Table~\ref{Results}.

   \begin{table}
   \caption{The masses and power-law indices for the CLUMPFIND,
    GAUSSCLUMPS and SExtractor samples}
   \label{Results}
   \begin{tabular}{lcccccc}
   \hline\hline
     &\#\tablefootmark{a}  & $M_{tot}       $&$M_{max} $      & $M_{min}$  &$M_{peak}$&$\alpha$\tablefootmark{b} \\ %
     &  & $[10^3\,M_\odot]$&$[M_\odot]$& $[M_\odot]$& $[M_\odot]$  &         \\%
    \hline
    & \multicolumn{6}{c}{ individual clump temperatures; lower cut-off limit: $50 M_\odot$} \\
    \hline
    CF  & 687  & 42.2 &4 652.2 & 5.2 &  15.8 & $1.95\pm 0.07$\\ %
    GC  & 371  & 42.6 &4 458.2 & 8.2 &  25.1 & $1.95\pm 0.06$\\ %
    SE  & 414  & 48.2 &2 972.8 & 5.7 &  15.8 & $1.89\pm 0.06$\\ %
    \hline
    & \multicolumn{6}{c}{T=10K; lower cut-off limit: $100 M_\odot$}\\
    \hline
    CF  &      & 69.1 &3 366.2 &14.7 &  39.8 & $2.15\pm 0.08$\\ %
    GC  &      & 69.9 &3 225.9 &14.6 &  63.1 & $2.08\pm 0.06$\\ %
    SE  &      & 78.8 &2 676.9 &15.4 &  39.8 & $1.99\pm 0.06$\\ %
    \hline
    & \multicolumn{6}{c}{T=20K; lower cut-off limit: $32 M_\odot$}\\
    \hline
    CF  &      & 21.0 &1 024.4 & 4.5 &  12.6 & $2.14\pm 0.08$\\ %
    GC  &      & 21.3 &  981.6 & 4.5 &  20.0 & $2.06\pm 0.06$\\ %
    SE  &      & 24.0 &  814.6 & 4.7 &  12.6 & $2.02\pm 0.06$\\ %
    \hline
   \end{tabular}
   \tablefoot{ The results are shown for the individual temperature
    case and the two constant temperatures. For each temperature we
    use an according conservative lower cut-off limit.}
   \tablefoottext{a}{Number of clumps}
   \tablefoottext{b}{power-law index; $dN/dM \propto M^{-\alpha}$}
   \end{table}

   \begin{figure}[h!]
   \centering
   \resizebox{\hsize}{!}{\includegraphics{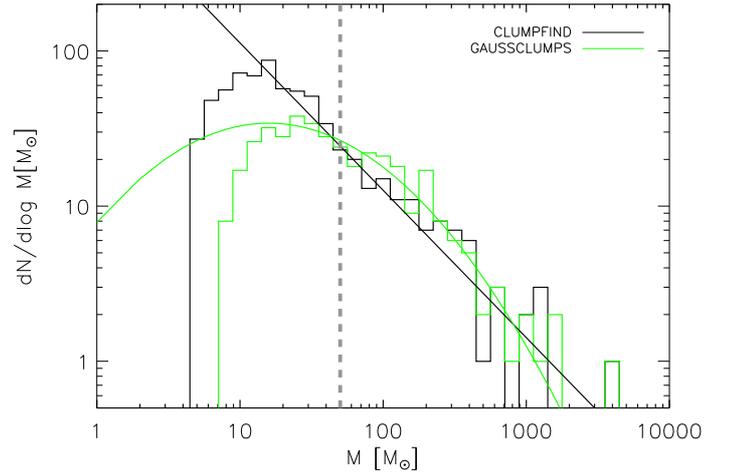}}
   \caption{
 The mass distribution of the CLUMPFIND sample with its power-law
 slope ($\alpha= 1.95$) in black. In green the GAUSSCLUMPS sample
 is over plotted with its log-normal fit. The dashed gray line shows
 the lower cut-off limit.}
   \label{log_norm}
   \end{figure}
   
\subsection{Power-law versus Log-normal}
\citet{Peretto10b} found, for a sample of gravitationally bound 
IRDCs, a power-law index $\alpha\sim1.8$, while the slope of the
mass spectra of unbound fragments steepens at the high mass end. They
also show that the mass function of the fragments can be well 
described by a log-normal distribution. These results are in 
agreement with the theoretical work of \citet{Hennebelle08} and with 
studies of the probability distribution function of the column 
density (N-PDF) within molecular clouds which displays  log-normal 
shapes for turbulent structures and power-law behaviour for 
gravitationally dominated clouds \citep{Kainulainen11, 
Ballesteros11}.

When we calculate the power-law slope for different lower cut-off 
limits, starting from the peak mass to higher masses, we find that 
the power-law index for our CLUMPFIND sample is quite robust, within 
the errors, regardless of whether we assume individual or constant 
temperatures for the clumps. The slopes of the other two samples 
however become continuously steeper while shifting the cut-off point 
to higher masses. This indicates that GAUSSCLUMPS and SExtractor 
samples can be also described by a log-normal function as well as by 
a power-law while the CLUMPFIND sample shows a clear power-law 
behavior (Fig.~\ref{log_norm}).
We thus find that, the results of the before mentioned studies have 
to be used with caution while drawing conclusions on the physical 
state of a single molecular cloud, as the shape of the ClMF can 
highly depend on the clump extracting method. Whether the clumps are
gravitationally bound or dominated by turbulence has to be confirmed
otherwise, e.g. by determining their virial parameter
\citep{Bertoldi92}. 

\section{Summary}
We analysed a large-scale sub-mm map of the Carina Nebula Complex to
investigate the molecular cloud structures of the region 
\citep{Preibisch11}. From this map we extract the clumps with three
common clump-finding algorithms (CLUMPFIND, GAUSSCLUMPS and 
SExtractor) to derive the ClMF. We therefore are able to compare the 
resulting ClMFs and to test the influence of the extraction method 
and the effects of the assumed temperatures on the ClMF (see Table~
\ref{Results}).

The assumption of a uniform temperature for the whole area disregards
that the different regions within the Carina Nebula Complex are
affected by highly variable levels of irradiation and stellar 
feedback. 
Hence we used an empirical calibration of the relation between column
density and temperature, which we adopted from data of 
\citet{Peretto10}, to estimate the temperatures and masses of the
individual clumps.
The clump temperatures for all extraction methods lie between
$8.5 - 18.5$~K, which are characteristic temperatures for molecular
clumps.

From the subsequent resulting masses we derived the ClMFs of this
region. For masses above $50\,M_\odot$ all samples can be described 
by a power-law with index $\alpha$ around 1.9, as defined in
Eq.~(\ref{power-law}). This is in good agreement with the results of  
other studies of clump mass distributions in molecular clouds 
\citep{Elmegreen96, Kramer98, Schneider04}. The slope of the
power-law we found is also quite similar to the slopes for cluster 
mass functions \citep[e.g.][]{Lada03}. If we assume a uniform 
temperature for all clumps (10~K / 20~K) the power-law slope gets 
steeper.

The GAUSSCLUMPS and SExtractor samples show a lack of clumps in the
low-mass range, while the CLUMPFIND algorithm finds for this mass 
regime at least twice the number of clumps. So even when the 
high mass end of the thus found ClMFs of the GAUSSCLUMPS and 
SExtractor samples show a power-law behaviour their overall shapes 
are better described by a log-normal function.
The CLUMPFIND sample, however, clearly follows a power-law. Hence we 
find the shape of the ClMF highly dependent on the method used to 
extract the clumps.

Theoretical models predict that unbound turbulent clouds should
have a log-normal ClMF, whereas clouds containing gravitationally
bound clumps develop a power-law tail \citep{Kainulainen11, 
Peretto10b}. In this sense, the shape of the ClMF contains very 
important information about the evolutionary state of a molecular 
cloud. However, our results show that a reliable determination of the
shape of the ClMF (log-normal or power-law) is often difficult as it 
can depend on the specific source extraction algorithm. Therefore, 
conclusions drawn from a ClMF derived from a single extraction 
algorithm should be taken with care to avoid an over-interpretation. 

\begin{acknowledgements}
This work was supported by funding from
Deutsche Forschungsgemeinschaft (DFG) under projectnumber PR 569/9-1.
Additional support came from funds from the 
Munich Cluster of Excellence: ``Origin and Structure of the
Universe''. S.~Pekruhl was financially supported from the 
International Max-Planck Research School on Astrophysics (IMPRS),
Garching. We also want to thank T.~Ratzka for many helpful 
discussions.
\end{acknowledgements}

\bibliographystyle{aa} 
\bibliography{bibliography} 

\end{document}